\documentclass[aip,apl,reprint]{revtex4-1}

\usepackage{graphicx}
\usepackage{amsmath}

\begin{document}


\title{Electrostatic tuning of magnetism at the conducting (111) (La$_{0.3}$Sr$_{0.7}$)(Al$_{0.65}$Ta$_{0.35})$/SrTiO$_3$ interface }

\author{V. V. Bal$^1$, Z. Huang$^{2,3}$, K. Han$^{2,3}$, Ariando$^{2,3,4}$, T. Venkatesan$^{2,3,4,5,6}$ and V. Chandrasekhar$^{*}$}
\affiliation{$^1$Department of Physics, Northwestern University, Evanston, IL 60208, USA,\\ 
 $^2$ NUSNNI-Nanocore, National University of Singapore 117411, Singapore\\
 $3$ Department of Physics, National University of Singapore 117551, Singapore\\
 $4$ NUS Graduate School for Integrative Sciences and Engineering, National University of Singapore 117456, Singapore\\
 $5$ Department of Electrical and Computer Engineering, National University of Singapore 117576, Singapore\\
 $6$ Department of Material Science and Engineering, National University of Singapore 117575, Singapore}

\date{\today}                              
                            
\begin{abstract}
We present measurements of the low temperature electrical transport properties of the two dimensional carrier gas that forms at the interface of $(111)$ (La$_{0.3}$Sr$_{0.7}$)(Al$_{0.65}$Ta$_{0.35}$)/SrTiO$_3$ (LSAT/STO) as a function of applied back gate voltage, $V_g$.  As is found in (111) LaAlO$_3$/SrTiO$_3$ interfaces, the low-field Hall coefficient is electron-like, but shows a sharp reduction in magnitude below $V_g \sim$ 20 V, indicating the presence of hole-like carriers in the system.  This same value of $V_g$ correlates approximately with the gate voltage below which the magnetoresistance evolves from nonhysteretic to hysteretic behavior at millikelvin temperatures, signaling the onset of magnetic order in the system.  We believe our results can provide insight into the mechanism of magnetism in SrTiO$_3$ based systems.
\end{abstract}

\maketitle
For more than ten years, the two dimensional carrier gas in SrTiO$_3$ (STO) based heterostructures has provided us with a model system to study a wide variety of physical phenomena, such as magnetism, superconductivity, spin-orbit interaction, and localization effects.\cite{Ohtomo,Caviglia,Thiel,Reyren,Brinkman,CavigliaSOC,Dikin}  Adding to the interest is the fact that these interfacial phenomena can be tuned by a range of experimental handles, including applied electric fields,\cite{Thiel} oxygen partial pressure during growth,\cite{Brinkman,Ariando,Siemons} capping layers,\cite{Lesne} post growth annealing treatment,\cite{Ariando2,Davis} crystal orientation,\cite{Herranz} and strain.\cite{CBEom}  So far, most studies on these heterostructures have focused on LaAlO$_3$ (LAO) deposited on a (001) oriented STO substrate.  However, it was recently discovered that changing the surface crystal orientation can vastly change the properties of the carrier gas.\cite{Herranz}  In particular, a tunable anisotropy in transport parameters was observed in (110) and (111) oriented LAO/STO.\cite{Davis,Davis2,Gopinadhan,Dagan,Herranz2}  The (111) oriented interface has also been predicted to show novel topological phases, given that the surface Ti atoms in STO form a honeycomb lattice that hosts orbitals with hexagonal symmetry.\cite{Syro,Walker,Pickett,Xiao,Yang}

\begin{figure}[tbp]
      \begin{center}
      \includegraphics[width=9cm,height=19cm]{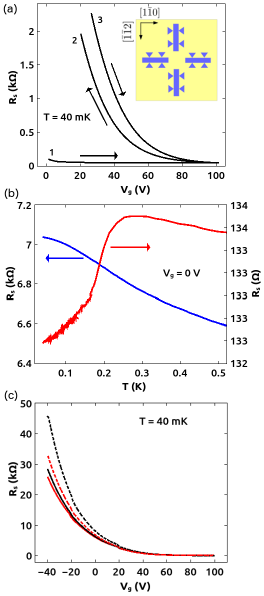}
      \caption{\textbf{(a)} Sheet resistance $R_s$ vs. $V_g$ at 40 mK, immediately after the initial cooldown at $V_g$ = 0 V.  Arrows indicate the direction of $V_g$ sweep, with the first $V_g$ sweep (trace 1) going from $V_g$ = 0 V to 100 V.  The inset shows the sample geometry, with the etched STO (yellow), and the two Hall bars (blue) oriented along [$\bar{1}\bar{1}$2] and two along [1$\bar{1}$0].  \textbf{(b)} $R_s$ vs. $T$ at $V_g$ = 0 V, at initial cooldown (on the left axis), and after gate sweeps (on the right axis), for one of the Hall bars.  \textbf{(c)} $R_s$ (averaged over up and down gate voltage sweeps) vs. $V_g$ of all four Hall bars after sweeping $V_g$ between -40 V and 100 V, at 40 mK, showing absence of obvious anisotropy between crystal axes.  The dashed and continuous traces in red are data for two Hall bars oriented along a single crystal axis, whereas the dashed and continuous traces in black are data for the two Hall bars oriented along the other crystal axis.}
      \label{fig1}
     \end{center}
\end{figure}

(La$_{0.3}$Sr$_{0.7}$)(Al$_{0.65}$Ta$_{0.35}$) (LSAT) is a commonly used substrate for the growth of other perovskite films.  When LSAT is grown on STO, a two-dimensional carrier gas forms at the LSAT/STO interface.\cite{Huang}  LSAT has a band gap of 4.9 eV,\cite{Nunley} smaller than the LAO band gap of 5.6 eV,\cite{Lim} which can affect the relative importance of the different mechanisms contributing to the formation of the conducting gas.  LSAT has a 1 \% lattice mismatch with STO, as opposed to a 3 \% mismatch in the case of LAO, which means the interfacial STO layers experience a smaller strain in the case of LSAT/STO.  Also, unlike LAO, but similar to STO, LSAT undergoes a cubic to tetragonal transition at about 100 K,\cite{Cowley,Chakaomakos} which is also expected to reduce strain effects.  Earlier work on (001) LAO/STO interfaces\cite{CBEom} demonstrated that strain at the interface can drastically change conduction, whereas recent studies of LSAT grown on (001) STO suggested that the lower strain compared to LAO significantly increased the carrier mobility.\cite{Huang,Huang2}  However, there have been only a few transport experiments performed on LSAT/STO interfaces to date, and none on (111) oriented LSAT/STO.  In particular, the effect of an electric field, which has proved to be a powerful tool to probe the properties of other STO based interface devices, has not been studied in the case of LSAT/STO.  

We report here measurements of the low temperature transport properties of (111) oriented LSAT/STO as a function of the back gate voltage $V_g$ and applied perpendicular magnetic field $B$.  As previously found with (001) LSAT/STO, we find that the devices show a very high residual resistance ratio (RRR) on the initial cooldown from room temperature, but the low temperature resistance of the devices increases significantly on applying a positive $V_g$.  This is similar to what has been observed in LAO/STO interfaces,\cite{Biscaras} where the change was attributed to an irreversible leakage of charge carriers from the interface to the bulk.  The Hall coefficient shows a sharp drop in magnitude below $V_g \sim 20$ V, indicating the presence of holes along with electrons at the interface, in agreement with what has previously been observed for the (111) LAO/STO interface.\cite{Davis}  The striking observation we make here is that at millikelvin temperatures, the magnetoresistance (MR) evolves from being nonhysteretic at high positive values of $V_g$ to strongly hysteretic at lower values of $V_g$, indicative of an electrically tuned ferromagnetic phase.  The gate voltage below which strongly hysteretic MR is observed corresponds approximately to the value of $V_g$ below which the Hall coefficient drops sharply in magnitude.  Electric field tuned magnetism has been observed before in LAO/STO by Bi \textit{et al},\cite{Bi} who associated the appearance of magnetism with strong charge depletion of the 2DEG, \textit{i.e.}, the emergence of an insulating system with localized magnetic impurities.  The (111) LSAT/STO interfaces in this study remain conducting even in the strongly hysteretic regime, pointing to a different origin of the magnetism in the system.\\

\begin{figure}[tbp]
      \begin{center}
      \includegraphics[width=8.5cm]{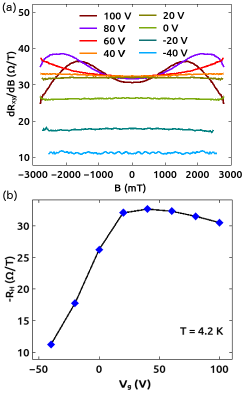}
      \caption{\textbf{(a)} Derivatives of transverse MR, $dR_{xy}/dB$, as a function of $B$, for various values of $V_g$.  Data were taken at 4.2 K. \textbf{(b)} Hall coefficient, defined as $dR_{xy}/dB$ at $B=0$, as a function of $V_g$, taken at 4.2 K.}
      \label{fig2}
     \end{center}
\end{figure}

The devices were fabricated on 12 monolayers of LSAT grown epitaxially by pulsed laser deposition (PLD) on (111) STO, at an oxygen partial pressure of 10$^{-4}$ Torr.  No post growth anneal was performed.  The details of the PLD synthesis have been discussed in earlier papers. \cite{Huang,Huang2}  Using photolithography and Ar ion milling, four Hall bars (each 600 $\mu$m long and 100 $\mu$m wide) were patterned onto one 5 mm x 5 mm chip, so that the lengths of two Hall bars were aligned along the [1$\bar{1}$0] surface crystal direction, and lengths of the other two Hall bars were aligned along the [$\bar{1}\bar{1}$2] crystal direction (see inset to Fig. \ref{fig1}(a)).  Ti/Au was deposited to define the contact pads and the devices were wire bonded with Al wires.  The chip was mounted with silver paint on an electrically contacted copper puck to enable application of a back gate voltage through the 0.5 mm thick STO substrate using a Keithley voltage source.  The measurements were carried out in a Kelvinox MX100 dilution fridge with a base temperature of 40 mK using standard 4-probe ac measurements at a frequency of 3 Hz.  The ac excitation current was $\sim$ 10 nA at millikelvin temperatures and $\sim$ 100 nA at 4.2 K.  Unlike in the case of (111) LAO/STO samples,\cite{Davis2} we did not observe any significant anisotropy in the electrical transport properties between Hall bars aligned along the two surface crystal directions (Fig. \ref{fig1}(c)).  Consequently, in what follows, we discuss in detail measurements for one of these Hall bars; the other three Hall bars gave results in qualitative agreement.   \\

\begin{figure}[tbp]
      \begin{center}
      \includegraphics[width=8.5cm]{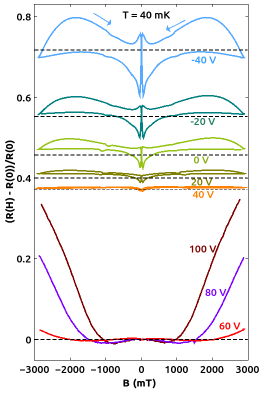}
      \caption{Fractional change in MR for various $V_g$.  The data are shifted along the vertical axis for clarity.  The dashed black lines are an aid to the eye indicating the vertical shift of the origin for the various values of $V_g$.}
      \label{fig3}
     \end{center}
\end{figure}

The devices were first cooled to base temperature with the back gate grounded.  The 4-probe $R_s$ was $\sim$10 k$\Omega$ at room temperature, and dropped to $\sim$0.1 k$\Omega$ at 4 K, giving a RRR$\sim$ 100. A large RRR implies that the sample is very clean, with a small amount of disorder.  However, after sweeping the back gate to positive voltages ($V_G$=100 V), we found that the resistance increased drastically and irreversibly at low temperatures (Fig. \ref{fig1}(a)).  Subsequent gate voltage sweeps retraced this higher resistance state over our entire gate voltage range with some hysteresis, as shown by traces 2 and 3 in Fig. \ref{fig1}(a).  The irreversibility also manifested itself as a change in the low temperature dependent resistance (Fig. \ref{fig1}(b)):  prior to applying a positive gate voltage, the resistance showed a drop in resistance below $\sim 250$ mK for $V_g$=0 V, a potential hint of a superconducting transition.  After sweeping to $V_g$=100 V, the low temperature resistance at $V_g$=0 V, in addition to increasing by a factor of $\sim$ 50, showed a characteristically different dependence, increasing with decreasing temperature.  The low temperature, low resistance behavior was only recovered on warming the sample to room temperature.  Similar behavior has been previously observed in (001) LAO/STO samples, and was attributed to an irreversible loss of carriers from the potential well at the interface to the bulk of the STO, brought about at high positive $V_g$ due to a change in the shape of the interfacial potential well.\cite{Biscaras}  For our LSAT/STO samples, measurements of the Hall resistance at any particular value of $V_g$, before and after sweeping $V_g$ at low temperatures showed little change, indicating that perhaps a large decrease in mobility of the charge carriers (rather than a drastic change in carrier density) on sweeping $V_g$ to positive values may be the primary reason for the increase in sample resistance. \\

Figure \ref{fig1}(c) shows the averaged low temperature longitudinal resistance as a function of $V_g$ for all four  Hall bars, two aligned along the [1$\bar{1}$0] direction, and the other two along the [$\bar{1}\bar{1}$2] direction.  Although the four curves are different due to inhomogeneities in the sample as is expected in two dimensional carrier gases based on STO, there is no evidence of systematic anisotropy based on crystalline direction.  This is in contrast to the behavior seen in (111) LAO/STO, where for samples with comparable $R_s$ to ours, the resistance measured along the [1$\bar{1}$0] direction can be a factor of 5 or more greater than the resistance along the [$\bar{1}\bar{1}$2] direction at large negative gate voltage.\cite{Davis2}  As noted above, one of the significant differences between the two systems is the greater strain at the LAO/STO interface, suggesting that this strain may be responsible for the observed anisotropy in (111) LAO/STO.  \\

Figure \ref{fig2}(a) shows the derivative of the Hall resistance $dR_{xy}/dB$ for various values of $V_g$ at 4.2 K.  For high positive values of $V_g$, we see nonlinearities in the Hall resistance at larger fields, which disappear as $V_g$ is reduced.  The nonlinear behaviour is indicative of multicarrier transport, as has been observed before in LAO/STO interfaces.\cite{Kim,Joshua} The value of $dR_{xy}/dB$ at $B=0$, which we define as the Hall coefficient $-R_H$, is shown in Fig. \ref{fig2}(b).  $-R_H$ varies slightly above $V_g=20$ V, but decreases rapidly below this gate voltage.  This behavior is similar to what is observed in (111) LAO/STO films.\cite{Davis}  If all the carriers are electron-like, one would expect a decrease in electron density $n_e$, and hence an increase in the magnitude of the Hall coefficient as $V_g$ is decreased, opposite to what is observed.  Consequently, we believe that holes, in addition to electrons must contribute to transport at the (111) LSAT/STO interface, as has already been shown for the (111) LAO/STO interface.\cite{Davis}\\
 
The most striking feature of the low temperature transport properties of these devices is found in the longitudinal MR, shown in Fig. \ref{fig3}.  The gate voltage $V_g\sim 20$ V below which $|R_H|$ shows a sharp drop correlates approximately with the gate voltage below which the longitudinal MR starts showing significant hysteresis, a signature of magnetic order in the system.  For $V_g \geq 60$ V, the longitudinal MR is not hysteretic:  at large magnetic fields,  it shows an approximately quadratic positive MR background that we identify with the classical MR.  The classical MR is important when $\omega_c \tau>1$, where $\omega_c = eB/m$ is the cyclotron frequency, $\tau$ the scattering time, and $m$ the carrier mass.\cite{Ashcroft}  The fact that this quadratic dependence starts at relatively low magnetic fields ($B\sim 1$T) means that $\tau$ is correspondingly large, \textit{i.e.}, the system is relatively clean, consistent with the low value of resistance seen in this gate voltage regime.  Near zero field, for $V_g \geq 60$ V, we see a small negative MR that we associate with weak localization, which will be studied in greater detail in future publications.  For $V_g=40$ V, we see a small hysteresis in the MR, which becomes progressively larger as $V_g$ is reduced.  We note that the hysteresis is observed only at millikelvin temperatures, and disappears when the MR is measured at 4.2 K.  Hysteresis in the longitudinal MR has been observed in (001) LAO/STO devices, and is understood as a signature of underlying ferromagnetic order,\cite{Dikin} the presence of which has been confirmed by other techniques as well.\cite{Moler,Ashoori}  The MR data from the (111) LSAT/STO devices is different in some important respects.  First, while hysteresis in the MR is observed over the entire measured gate voltage range for (001) LAO/STO,\cite{Mehta} in the (111) LSAT/STO devices discussed here, it appears only for $V_g \leq 40$ V, correlating approximately with the voltage below which $|R_H|$ shows a sharp drop.  Second, the hysteresis in the MR observed in (001) LAO/STO samples occurs over a very narrow range of field near $B$=0, and has been associated with the coercive field of the underlying ferromagnet.  At larger field scales, the MR is not hysteretic.  In contrast, the hysteresis in the MR in our (111) LSAT/STO devices at negative $V_g$ is present over the entire range of our magnetic field.  The correlation of the onset of the hysteretic MR with the sharp drop in $|R_H|$ suggests that the magnetism is associated with population/depopulation of specific electronic bands at the interface.  \\

In summary, we have measured the low temperature electronic transport properties of (111) oriented LSAT/STO interfaces as a function of an applied back gate voltage $V_g$.  We find that the MR becomes hysteretic below a specific value of $V_g$ that corresponds approximately to the voltage below which the Hall coefficient shows a sharp drop in magnitude.  The hysteretic MR is evidence of the emergence of a gate-tunable magnetism.  Further experimental and theoretical work is required to understand the origin of this magnetism.\\

The U.S. Department of Energy, Office of Basic Energy Sciences supported the work at Northwestern University through Grant No. DE-FG02-06ER46346.  Work at NUS was supported by the MOE Tier 1 (Grants No. R-144-000-364-112 and No.R-144-000-346-112) and Singapore National Research Foundation (NRF) under the Competitive Research Programs (CRP Awards No. NRF-CRP8-2011-06, No. NRF-CRP10-2012-02, and No. NRF-CRP15-2015-01).  This work utilized Northwestern University Micro/Nano Fabrication Facility (NUFAB), which is supported by the State of Illinois and Northwestern University.

$^*$v-chandrasekhar@northwestern.edu

\end{document}